# Grid-Aware Provision and Activation of Fast and Slow Flexibilities from Distributed Resources in Low and Medium Voltage Grids

Ankur Majumdar, *Member, IEEE*, Omid Alizadeh-Mousavi

*Abstract*—As more and more renewable intermittent generations are being connected to the distribution grid, the grid operators require flexibility to maintain the balance between supply and demand. The intermittencies give rise to situations which require not only slow-ramping flexibility capability but also, fast-ramping flexibility capability from a variety of resources connected at the MV and LV distribution grids. Moreover, the intermittencies may increase the costs of grid reinforcement. Therefore, to defer the reinforcement of the grid assets, the grid needs to be operated optimally. This paper proposes – a) such an optimal operational methodology for the MV and LV grids; and b) an aggregated flexibility estimation methodology estimated separately for fast and slow services at the primary substation (TSO interface). The methodologies based on model-based MV grids and a sensitivity coefficients-based model-less LV grids are suitable for LV grids where an up-to-date and accurate model and topology are not always available. The approaches of the paper use the synchronised and accurate measurements from grid monitoring devices located at the LV distribution grids. It is assumed that the implementation of the methodology is centralised, where a grid monitoring device or a central platform is capable to host grid aware algorithms and to communicate control setpoints to DERs. The approaches have been validated on a real MV and LV networks of a Swiss DSO equipped with grid monitoring devices. The results, in terms of technical losses, grid violation costs and flexibility capability curve, show the efficacy of the optimal operation and flexibility estimation methodologies and therefore, can be easily deployed.

## Nomenclature

| | |
|---|---|
| DSO | Distributed system operator |
| TSO | Transmission system operator |
| P-SS | Primary substation |
| S-SS | Secondary substation |

## I. Introduction

THE modern distribution grid is gradually transitioning from a passive to an active system with increased number of distributed energy resources (DERs) being connected to the

This work is supported by the European Union's Horizon 2020 programme under the Marie Skłodowska-Curie grant agreement no. 840461.

A. Majumdar and O. Alizadeh-Mousavi are with Depsys SA, Route du Verney 20B, Puidoux, Switzerland (e-mail: ank.majumdar@gmail.com; omid.mousavi@depsys.ch).

medium voltage (MV) and low voltage (LV) grids. Moreover, a large number of DERs' outputs are intermittent in nature depending on available sunshine (solar photovoltaics) or available wind (wind farms). This means there is an increasing challenge to maintain production and consumption balance. In this scenario, the distribution system operators (DSOs) need to play a more proactive role by having provision of and ability to activate a larger portfolio of flexible resources connected at the LV level such as, roof-top PVs, electric vehicles (EV), energy storage systems (ESS), interruptible loads [1].

### A. Monitoring for flexibility provision and activation

To achieve this, the DSOs across the world are going through a process of digitalisation of their grids [2],[3]. Due to the cost of faster data transfer and storage and data privacy issues associated with the smart meter data, it renders them difficult to be used for real-time operational decisions such as, provision and activation of flexibility. The micro phasor measurement units (µPMUs) installed at the MV distribution grids come at a very high cost due to the requirement of high accuracy for angle measurements and the requirement of high data communication and data storage in the form of phasor data concentrators [4]. At this juncture, grid monitoring devices at the LV grid have the capability to provide synchronised, near real-time and accurate data on voltages, line currents, active and reactive power flows for flexibility provision and activation without the above-listed difficulties of µPMUs and smart meters [5]. Moreover, they are beneficial for optimal operation of the MV grid [6] and can communicate control setpoints to small-scale flexible DERs connected to the LV grid. This means that the monitoring and control functionalities of LV monitoring technology can facilitate flexibility provision available from the LV grid and therefore, not only improve the security and reliability of operation of the distribution grid but also, enable optimal utilisation of grid assets (lines, transformers etc.) and flexible DERs.

### B. Background review on use of flexibilities

Over the years, there has been significant focus on optimal operation of active distribution grids. This enables not only to ensure the security, quality and continuity of supply but also, efficiently manage grid resources and assets. The optimal operation enables the DSOs to defer the grid reinforcement of distribution lines and transformers through non-network solutions of active network management [7],[8].

As the regulators have set ambitious targets of decarbonising the electricity system, both TSO and DSOs, with increased number of intermittent resources, are faced with the challenge to operate their grid in situations which require not only slow-ramping services but also, fast-ramping services from a variety of resources [9]. To increase the liquidity or portfolio of these ancillary services, the modification of national regulations around the world allows the TSO to procure ancillary services (both fast and slow) from DERs connected at the distribution grid as well, while maintaining the security of the distribution grid [10]. Therefore, there is an increasing requirement for improved coordination in terms of data or information between the TSO and the DSOs. In [11],[12] several coordination schemes have been proposed, where the priority is given to either the TSO or the DSO. It is important that both the DSO and TSO grids remain secure in all the coordination schemes.

On the non-regulated side, to avail the opportunity of having increased number of flexible resources, there are several players in the market such as, aggregator, virtual power plants (VPP), suppliers with demand response, who can provide a wide range of services for the utilities (TSO and DSO) – from **fast** frequency responses such as, primary response, ramping services, to **medium** and **slow** such as, secondary and tertiary responses; and voltage control and congestion management services; and peak-shaving and load levelling services. In the current scenario, the entities such as, demand aggregator units, aggregators and VPPs provide flexibility services (fast and slow) from aggregated resources such as, DERs, EVs, ESS, thermostatically controlled loads (TCL), direct load control connected at the MV and LV levels [12-16]. However, the operational objectives of each of these aggregated bodies is different. While the demand aggregated units work on reducing the cost of energy consumption for the end-customers, the aggregator and VPPs work on a profit model. However, the full flexibility potential of the DERs is not realised due to distribution grid security limitations. Moreover, the aggregators, VPPs do not ensure optimal utilisation of distribution grid assets and optimal grid operation. Hence, there is a motivation from the DSO point of view for a grid aware algorithm to ensure an optimal operation of the grid and to facilitate the provision of fast and slow flexibility services.

*C. Research gaps*

The services offered from distributed resources and aggregated entities vary from country to country [17]. Depending on the mechanisms or remunerations available to participate in the market, a resource/aggregation of resources can be utilised for single or multiple services [18]. It is argued that energy storage systems, for example, can have higher returns if utilised for multi-service applications [19]. Even though the contract for providing flexibility to the TSO can be either committed or flexible, a large portion of resources connected at the DSO level has flexible contracts [20]. Indeed, an aggregator or VPP with its aggregate of resources can participate in fast and slow services depending on the controllability in terms of response time and duration and availability of its resources [17],[21]. However, to the best of knowledge of the authors, none of the literature addressed the problem of provision of fast and slow flexibility services from the distribution grid whilst ensuring the optimal operation and security of the grid, with model-less LV grid, through grid aware algorithms.

*D. Innovative aspects*

The work presented in this paper has the following innovative aspects.

1) Firstly, it proposes a grid-aware approach for provision and activation of flexibility considering the physical constraints of MV and LV grids together. The combined methodology considers the full model of topology and parameters of the MV grid and a model-less approach for the LV grid.

2) Secondly, the model-less approach for the LV grid based on sensitivity coefficients does not require an up-to-date, accurate and trustable information on the LV grid topology and parameters. As the access to this information is not always available to the DSO, this methodology is particularly useful for achieving an optimal operation of the whole distribution grid. Besides, this approach is robust to changes in the grid topology and operating points of the LV grid.

3) Thirdly, this paper proposes a methodology of estimation and activation of fast and slow flexibility services separately from resources connected at the MV and LV levels. The idea is based on estimating fast and slow flexibility services at the primary substation (P-SS) level with grid aware approach (combined MV-LV OPF); and then activating them when required.

4) Fourthly, the grid aware algorithms (centralised implementation) are either hosted by a grid monitoring device or a DSO central platform. This plays a crucial role in managing this estimation and activation of fast and slow services. It is also crucial to consider the coordination scheme between the TSO and the DSO to better facilitate this activity. It is assumed, here, that the aggregated flexibility (fast or slow) is utilised by the TSO once the DSO pre-qualifies the services considering that they do not cause congestion or voltage problems in the distribution grid.

The remainder of the paper is structured as follows. Section II discusses the definition of fast and slow flexibility services and the coordination between the TSO and the DSO. Section III describes the mathematical formulations of the proposed methodologies for grid-aware estimation and activation of flexibility services. Section IV illustrates the results and discusses the insights and outcomes of the results. Section V, the last section, concludes the paper.

## II. FAST AND SLOW FLEXIBILITY SERVICES

*A. Definition and classification of services*

A flexibility service is defined as fast or slow depending on how quickly (response time) a resource can respond to requests from the grid operator or for a change in setpoint. Therefore, a service is classified as fast or slow according to the ramp rate of active or reactive powers. The grid operators require flexibility services for maintaining the grid frequency within operational and statutory limits, node voltages and power flows within operational limits. Some flexible resources can also be utilised for black start capability and balancing capacity reserves. The fast and slow services, with the standard names of the products, required by the TSO and the DSOs are summarised in Table I.

TABLE I
FAST AND SLOW FLEXIBILITY SERVICES

|  | Fast services | Slow services |
|---|---|---|
| Required by TSO | • Inertia (milli-seconds, not available now)<br>• Primary frequency response (FCR, FFR, EFR)<br>• Secondary reserves (aFRR)<br>• Ramping response | • Tertiary response (mFRR)<br>• Tertiary response (RR)<br>• Voltage and reactive power<br>• Congestion<br>• Black start<br>• Balancing capacity reserves |
| Required by DSO | No standard service available now | • Voltage<br>• Congestion<br>• Peak shaving<br>• Load levelling |

Although the classification of fast and slow services remains largely the same, as shown in Table I, the names of the standard and specific products can vary from one regulatory region to the other [22-24]. It is argued that primary frequency response is mainly provided by transmission connected resources such as, large generators, large wind farms and grid energy storages. Primary response can also be provided by distribution connected DERs however, there are minimum capacity limitations for providing the same. It is unlikely that these resources connected at the LV grid can participate in the primary response and therefore, is outside the scope of the paper. The very fast inertial response is provided by large generators and there are no market-based services available, thus, it is outside the scope of this paper. A secondary response usually must be activated within 5 seconds and may last for 15 minutes. A tertiary response can be a back-up or replacement for the secondary response and must be available within 15 minutes. The voltage control and congestion management services are much slower than frequency and can be cleared every 30 minutes. The black start and balancing capacity are usually capacity reserves for unforeseen circumstances. As for the DSO, there are no fast services defined till now. The DSO slow services can range from voltage control, congestion management to peak shaving. In this work, the fast and slow flexibility services available from distributed resources in MV and LV grid will be communicated through separate feasible flexibility curves considering the security aspects of the distribution grid. In the case when either fast or slow service is activated, it is important to ensure that activation of either service will not affect the security of the distribution grid.

### B. TSO-DSO coordination schemes

The provision and activation of fast or slow services depend on the data or information coordination scheme between the TSO and the DSO. According to the coordination schemes discussed in [11],[25], they can be broadly classified into four schemes – i) centralised common TSO-DSO market, ii) single ancillary services market (TSO leader), iii) local ancillary services market (DSO leader), iv) shared balancing responsibility. It is argued that a common market platform for TSO and DSOs, being computationally expensive, may not be practically feasible. In the shared balancing responsibility scheme, each DSO runs a local market platform with the TSO having no access to DSO-level resources. However, with regulations advocating the use of distribution-connected resources for TSO ancillary services, it is highly unlikely that this scheme will be taken up by the grid operators. Among the other two, they can be classified into either a scheme where the DSO takes the lead (local ancillary services) or where the TSO takes the lead (single ancillary services).

In the <u>DSO leader scheme</u> for the TSO-DSO coordination.

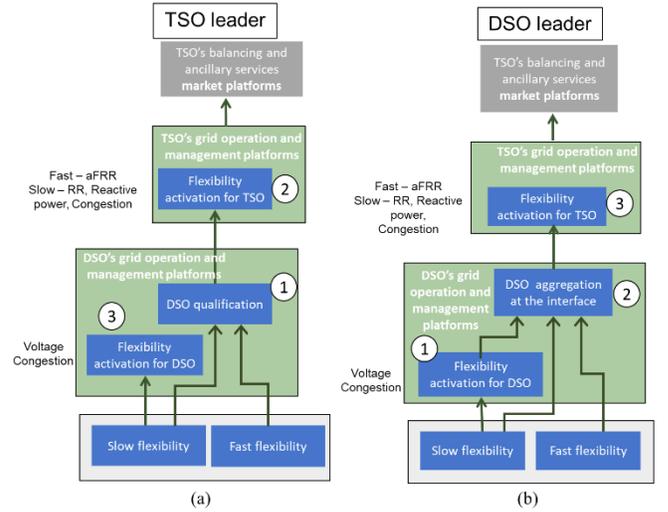

Fig. 1. TSO-DSO coordination schemes with (a) TSO leader and (b) DSO leader

i.e. the DSO has priority for contracting flexibility connected to the distribution grid. The TSO can contract flexibilities from the distribution grid with the aggregated flexibility at the TSO-DSO interface, only after the DSO has utilised the distributed flexibilities to maintain its own grid security. The DSO market clears before the TSO ancillary services market. The DSO leader scheme, however, can function both with and without, a market platform for the DSO. For the <u>TSO leader scheme</u>, the TSO contracts flexibilities from the distribution grid. Nevertheless, there is a pre-qualification phase supervised by the DSO, so that the distribution grid constraints are not violated. The TSO leader scheme is in line with the current practice of managing flexibility services by the grid operators. The distribution grid operation with the model-based MV grid and model-less LV grid, and the provision of separate fast and slow flexibility services ensure that the services are pre-qualified in the TSO leader scheme in relation to grid security. Moreover, the proposed methodology can also ensure the DSO utilises the flexibilities before the aggregation at the P-SS in the DSO leader scheme. Therefore, the innovative aspects, proposed in this paper, can be easily deployed for both the coordination schemes. Hence, the TSO and the DSO leader schemes, as illustrated in Fig. 1, have been adopted in the paper. In Fig. 1(a) and 1(b), the numbers 1, 2 and 3 denote the order of the processes in the coordination schemes.

## III. Problem Formulations

### A. Combined optimal operation of MV and LV grids

The distribution grid has limited monitoring at the feeder-outs of the primary substations (HV/MV transformers). The LV grid of the distribution grid remains largely unmonitored. As a result, the flexibility potential of decentralised resources connected at the LV grid remains untapped. The operation of such a grid is decided by load allocation based on historical load profile and rating of the MV/LV transformers. This cause sub-optimal grid operation or grid security violation, e.g. voltage violation and congestion. However, the digitalisation of the distribution grid will usher in changes in the distribution grid operation. The optimal utilisation of resources allows deferral of reinforcement of network assets (transformer, lines). This will enable not only to realise the flexibility potential of the DERs but also increase the grid hosting capacity of intermittent renewable generations. The authors in earlier publications discussed about a stage-wise operation, which consists of two separate OPFs for MV and LV grids, where the active and/or reactive power control variables at the MV side of MV/LV transformers are constrained by the LV grid flexibility curves [26],[27]. Although this result has satisfactory performance in terms of grid security, the optimality of the overall grid operation is not ensured. Furthermore, this stage-wise operational methodology raises issues regarding the optimal operation of the distribution grid (MV & LV) when two or more transformers from the same secondary substation (S-SS) are supplying different loads. Therefore, to find the optimal operating point, this paper proposes an approach where, the optimal setpoints are decided based on the optimal power flow of MV and LV grids together.

Fig. 2 illustrates the difference of the combined approach from the separate approach of grid operation for the MV and the LV grids. In the combined approach, the LV grids are modelled by a linear sensitivity coefficients approach whereas, the MV grid is modelled by convexified second-order conic formulation. The grid operation setpoints are decided by a central operation and management platform. It is assumed that grid monitoring devices at the LV grid provide grid observability and there is infrastructure in place to communicate the setpoints from the central platform to DERs and DERs are capable to receive the control setpoints. The formulation of the combined OPF is described below.

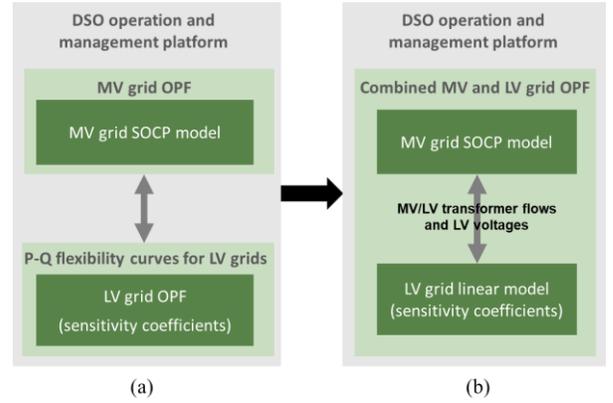

Fig. 2.(a) Separate OPFs and (b) combined OPF for MV and LV grids

$$P_{ij}^2(t) + Q_{ij}^2(t) \leq v_i(t) l_{ij}(t) \tag{5}$$

$$V_{dev}(t) = 0 \ \ if \ \ V^{min} \leq V_i(t) \leq V^{max} \tag{6}$$

$$V_{dev}(t) = |V_i(t)|^2 - V_{lim}^2, if \ V_i(t) < V^{min} \ or \ V_i(t) > V^{max} \tag{7}$$

$$\left|l_{ij}(t)\right|_{dev}^2 = 0 \ \ if \ |l_{ij}(t)| \leq I_{max} \tag{8}$$

$$\left|l_{ij}(t)\right|_{dev}^2 = \left|l_{ij}(t)\right|^2 - \left|l_{ij}\right|_{max}^2 \ if \ |l_{ij}(t)| > I_{max} \tag{9}$$

$$\Delta V_i^{LV}(\text{t}) = \textstyle\sum_k K_{VP,ik} * (\Delta P_k^{LV}(t)) + K_{VQ,ik} * (\Delta Q_k^{LV}(t)) \tag{10}$$

$$\Delta I_{ij}^{LV}(\text{t}) = \textstyle\sum_k K_{IP,ijk} * (\Delta P_k^{LV}(t)) + K_{IQ,ijk} * (\Delta Q_k^{LV}(t)) \tag{11}$$

$$V_i^{LV}(\text{t}) = V_{i,0}^{LV}(\text{t}) + \Delta V_i^{LV}(\text{t}) \tag{12}$$

$$I_{ij}^{LV}(\text{t}) = I_{ij,0}^{LV}(\text{t}) + \Delta I_{ij}^{LV}(\text{t}) \tag{13}$$

$$V^{min} \leq V_i^{LV}(\text{t}) \leq V^{max} \tag{14}$$

$$\left|l_{ij}(t)\right|^{LV} \leq I_{max} \tag{15}$$

$$V_{i,0}^{LV}(t) = 0.5(v_{i=MV}+1) - V_i^{dropLV}(t) \ \ \forall LV \ grids \tag{16}$$

$$-\left(p_k^g(t) + \Delta P_k^{LV}(t)\right) * \tan\left(\cos^{-1}0.95\right) \leq q_k(t) + \Delta Q_k^{LV}(t)$$
$$\leq \left(p_k^g(t) + \Delta P_k^{LV}(t)\right) * \tan\left(\cos^{-1}0.95\right) \tag{17}$$

$$(q_k + \Delta Q_k^{LV}(t))^2 + \left(p_k^g + \Delta P_k^{LV}(t)\right)^2 \leq 1.1 * s_{k,kVA}^2 \tag{18}$$

$$SOC_k(t) = SOC_k(t-1) + \eta\left(p_k^{EV}(t) + \Delta P_{k,EV}^{LV}(t)\right) \tag{19}$$
$$\forall \ t$$

The multi-objective framework (Eq. (1)) has the objective of reducing the network technical losses for the MV grid, reducing the security violation (voltage and congestion) costs and the costs of penalty on deviations in active and reactive schedules with the TSO over a time duration with each time step defined as $t$.

| Control variables | State variables | Parameters | |
|---|---|---|---|
| $\Delta P_k^{LV}$, $\Delta Q_k^{LV}$, $v_i$ ($i$ = slack) | $v_i$, $l_{ij}$, $P_{ij}$, $Q_{ij}$, $\Delta V_i^{LV}$, $\Delta I_{ij}^{LV}$, $V_i^{dropLV}$ | $K_{VP,ik}$, $K_{IP,ijk}$, $V_{i,0}^{LV}$, $I_{ij,0}^{LV}$ | $K_{VQ,ik}$, $K_{IQ,ijk}$, |

The variables and parameters of the optimisation problem are given in Table II. The formulation is constrained by power flow constraints (2)-(5). The Eqs (2)-(5) are the DistFlow equations for the MV grid, with $v_i(t) = |V_i(t)|^2$ and $l_{ij}(t) = \left|I_{ij}(t)\right|^2$, and convexified second-order conic formulation [28]. Eqs (6)-(9) represent the security constraints in terms of

$$min \ \textstyle\sum_t[w_I \sum_{lines} r_{ij}l_{ij}(t) + w_V \sum_{nodes} V_{dev}(t) + \\ w_{Ilim} \sum_{lines} l_{ij}{}_{dev}(t) + w_p p_{sl}^{MV}(t) + \\ w_q q_{sl}^{MV}(t)] \tag{1}$$

s.to

$$\textstyle\sum_{adj} P_{ij}(t) + \sum_{tr}(p_{sl}^{LV}(t) + \Delta p_{sl}^{LV}(t)) = p_i^g(t) - p_i^c(t) \tag{2}$$

$$\textstyle\sum_{adj} Q_{ij}(t) + \sum_{tr}(q_{sl}^{LV}(t) + \Delta q_{sl}^{LV}(t)) = q_i^g(t) - q_i^c(t) \tag{3}$$

$$v_j(t) = v_i(t) - 2r_{ij}P_{ij}(t) - 2x_{ij}Q_{ij}(t) + (r_{ij}^2 + x_{ij}^2)l_{ij}(t) \tag{4}$$

nodal voltages and line flows, which are piecewise linearised during the implementation. Eqs (10)-(15) denote the linearised power flow constraints for the LV grids, based on voltage (for node $i$) and current (for line $ij$) sensitivity coefficients $K_{VP,ik}$, $K_{VQ,ik}$, $K_{IP,ijk}$, $K_{IQ,ijk}$ with respect to active and reactive power change in node $k$, calculated around the operating points of node voltage $V_{i,0}^{LV}$ and branch current $I_{ij,0}^{LV}$. The DER inverters are assumed to operate with power factor control. Eq. (17) represents the power factor control for the inverters with a limit of 0.95 lead/lag power factor. Eq. (18) considers that the inverter can inject/absorb reactive power with an over-rated inverter. Eq. (19) represents the time-coupling state of charge constraints for the EVs and ESSs. The flows in the MV/LV transformers and voltage drop relationship for LV nodes with the MV and LV nodes of MV/LV transformer in the formulation ensure the logical connection between the MV grid and LV grid models, as shown in Fig 2(b). The MV grid formulation is related to the LV grid formulation with the active and reactive power flows through the MV/LV transformers $p_{sl}^{LV}$ and $q_{sl}^{LV}$ respectively, shown by Eqs (2)-(3). $\Delta p_{sl}^{LV}$ and $\Delta q_{sl}^{LV}$ represent the changes of active and reactive flows through the MV/LV transformers corresponding to the changes (flexibilities) in nodal injections of active and reactive power, respectively. Eq. (16) shows the relationship of the LV node voltages with the voltage at the MV/LV transformer node for each LV grid, where $V_i^{dropLV}$ is the drop in voltage with respect to the voltage at the primary of the MV/LV transformers. Eq. (16) is a linearisation of a non-linear equation $V_i^{LV}(t) = \sqrt{v_{i=MV}^{LV}(t)} - V_i^{dropLV}(t) + \Delta V_i^{LV}(t)$. The linearisation by Eq. (16) has negligible error, as shown in Fig. 3, when linearised around 1 p.u.

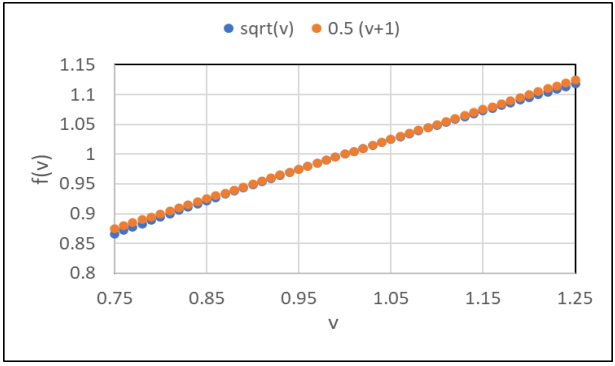

Fig. 3. Linearisation of square root of MV side variables

Fig. 4 illustrates the modification of the distribution grid with the detailed LV grid to a distribution grid with reduced linear sensitivity coefficients-based model-less LV grids and a MV grid with a full model. It further shows the combined MV and LV grid operational platform for the DSO with flexibility potential.

### B. Aggregated flexibility estimation for fast or slow services at P-SS

The LV grids of distribution grids have several controllable distributed energy production and consumption resources. Therefore, the LV grids have a wide range of flexible resources from fast to slow services. This means if needed, the LV grids are capable to provide both fast and slow services to the DSO and the TSO. As the TSO requests more flexibility services from the DSOs, the DSO must be capable to provide these services to the TSO at the same time ensuring the security and quality of supply of its own grid. This formulation describes an approach for the separate estimation of fast and slow flexibility services from the distribution grid at a primary substation. The formulation is based on the combined OPF explained in the previous subsection. The detailed formulation is given below.

$$max \ \sum_t [\alpha \, \Delta p_{sl}^{MV}(t) + \beta \, \Delta q_{sl}^{MV}(t)] \quad (20)$$

s.to

$$\sum_{adj} P_{ij}(t) + \sum_{tr}(p_{sl}^{LV}(t) + \Delta p_{sl,f/s}^{LV}(t)) = p_i^g(t) - p_i^c(t) \quad (21)$$

$$\sum_{adj} Q_{ij}(t) + \sum_{tr}(q_{sl}^{LV}(t) + \Delta q_{sl,f/s}^{LV}(t)) = q_i^g(t) - q_i^c(t) \quad (22)$$

$$v_j(t) = v_i(t) - 2r_{ij}P_{ij}(t) - 2x_{ij}Q_{ij}(t) + (r_{ij}^2 + x_{ij}^2)l_{ij}(t) \quad (23)$$

$$P_{ij}^2(t) + Q_{ij}^2(t) \le v_i(t)l_{ij}(t) \quad (24)$$

$$V^{min} \le V_i(t) \le V^{max} \quad (25)$$

$$|I_{ij}(t)| \le I_{max} \quad (26)$$

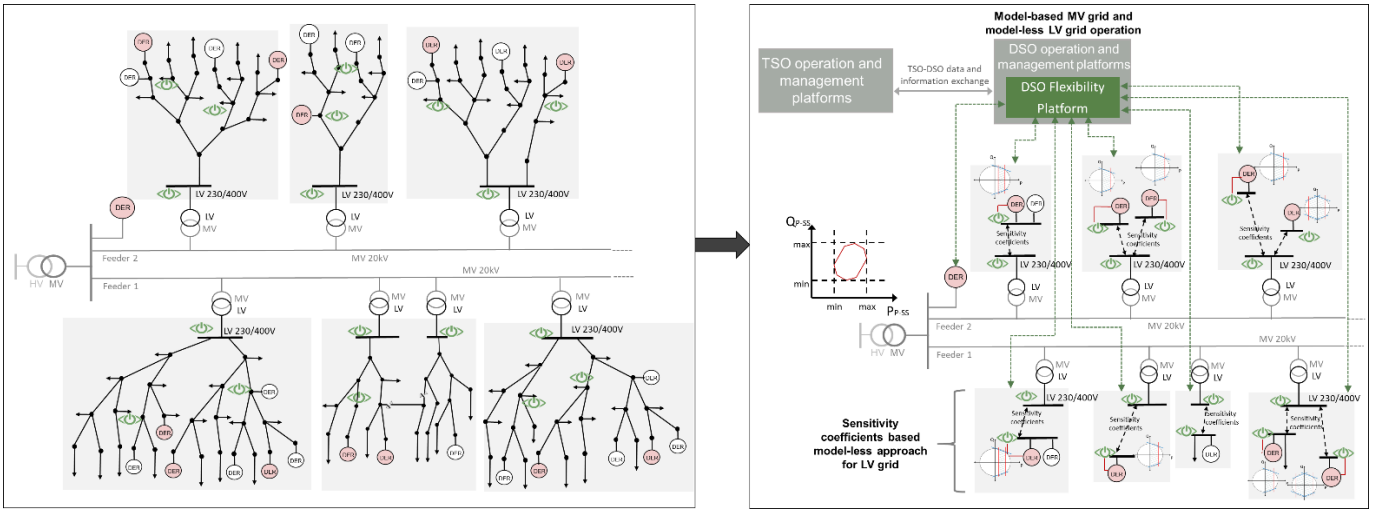

Fig. 4. A model-based MV grid and sensitivity coefficients-based model-less LV grid

$$\Delta V_i^{LV}(t) = \sum_k K_{VP,ik} * (\Delta P_{k,f/s}^{LV}(t)) + K_{VQ,ik} * (\Delta Q_{k,f/s}^{LV}(t)) \tag{27}$$

$$\Delta I_{ij}^{LV}(t) = \sum_k K_{IP,ijk} * (\Delta P_{k,f/s}^{LV}(t)) + K_{IQ,ijk} * (\Delta Q_{k,f/s}^{LV}(t)) \tag{28}$$

$$V_i^{LV}(t) = V_{i,0}^{LV}(t) + \Delta V_i^{LV}(t) \tag{29}$$

$$I_{ij}^{LV}(t) = I_{ij,0}^{LV}(t) + \Delta I_{ij}^{LV}(t) \tag{30}$$

$$V^{min} \le V_i^{LV}(t) \le V^{max} \tag{31}$$

$$\left| I_{ij}(t) \right|^{LV} \le I_{max} \tag{32}$$

$$V_{i,0}^{LV}(t) = 0.5(v_{i=MV}(t)+1) - V_i^{dropLV}(t) \quad \forall LV\ grids \tag{33}$$

$$-\left(p_k^g(t) + \Delta P_{k,f/s}^{LV}(t)\right) * \tan\left(\cos^{-1} 0.95\right)$$
$$\le q_k(t) + \Delta Q_{k,f/s}^{LV}(t)$$
$$\le \left(p_k^g(t) + \Delta P_{k,f/s}^{LV}(t)\right) * \tan\left(\cos^{-1} 0.95\right) \tag{34}$$

$$(q_k + \Delta Q_{k,f/s}^{LV}(t))^2 + \left(p_k^g + \Delta P_{k,f/s}^{LV}(t)\right)^2 \le 1.1 * s_{k,kVA}^2 \tag{35}$$

$$\Delta P^l \le \Delta P_{k,f/s}^{LV}(t) \le \Delta P^u \tag{36}$$

$$\Delta Q^l \le \Delta Q_{k,f/s}^{LV}(t) \le \Delta Q^u \tag{37}$$

$$R_k^{L,f/s} \le \left| \Delta P_{k,f/s}^{LV}(t) - \Delta P_{k,f/s}^{LV}(t-1) \right| \le R_k^{U,f/s} \tag{38}$$

$$R_k^{L,f/s} \le \left| \Delta Q_{k,f/s}^{LV}(t) - \Delta Q_{k,f/s}^{LV}(t-1) \right| \le R_k^{U,f/s} \tag{39}$$

$$SOC_k(t) = SOC_k(t-1) + \eta\left(p_k^{EV}(t) + \Delta P_{k,f/sEV}^{LV}(t)\right) \tag{40}$$
$$\forall t$$

The above multi-objective formulation aims to maximise the active and/or reactive power flows through the P-SS (Eq. (20)) for a time duration with each time step defined as $t$. The values α and β can take values in the range of [-1, 1]. For example, when α=1, β=0 to maximise the active power flow [29],[30]. The complete P-Q flexibility area is achieved by running the above formulation several times according to the required granularity of the area. As explained in the Subsection IIIA, Eqs (21)-(26) represent the power flow constraints and security constraints. The LV power flow equations are denoted by Eqs. (27)-(33). Similar to the previous subsection, the MV grid power flow is modelled by convexified DistFlow equations and the LV grids are modelled by linear sensitivity coefficients. $\Delta P_{k,f/s}^{LV}(t)$ and $\Delta Q_{k,f/s}^{LV}(t)$ are the changes in active and reactive powers, respectively, around the operating points. The subscript $f/s$ represents either fast or slow services. Eqs. (34)-(35) represent the power factor control and over-rating constraints for the DER inverters. Eqs (36)-(37) describe the upper and lower limits for the fast/slow active and reactive powers. The fast and slow services are decided based on the ramping limit (lower and upper) constraints (rate of change of active/reactive power), denoted by Eqs (38)-(39). Furthermore, the EVs and ESSs have time-coupling state of charge constraints given by Eq. (40).

## IV. RESULTS AND DISCUSSIONS

The proposed methodology has been tested and validated in a real network of a Swiss DSO with several GridEye grid monitoring devices at the secondary of the MV/LV transformers and the LV cabinets.

Fig. 5 shows the MV grid with several GridEye monitoring equipment at the LV side of the S-SS. Fig. 6 shows a representative LV grid with grid monitoring devices located at the LV side of the transformer and at the LV cabinets. The grid monitoring devices provide accurate synchronised 10-min measurement data of voltage, current, active and reactive powers. The test network has a total of 200kWp of solar PVs and 1000kWh of EV charging stations installed at the LV grids. It is assumed that the PVs have flexibility potential (negative) of up to 10% curtailment and EVs have flexibility potential (positive and negative) limited by charging/discharging limits of 8kW and state-of-charge limits from 0.1 to 0.9. The PV and EV aggregators inverters are assumed to operate with power factor control, with a limit of 0.90 power factor. The solar irradiance and power output of the PVs have been taken from renewables.ninja web tool [31]. The EV power output profiles have been provided by grid monitoring equipment installations at the EV charging stations. From the EV power outputs, a total of 12 EV profiles have been generated.

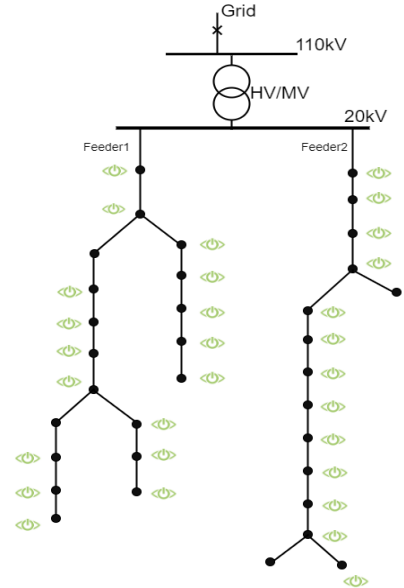

Fig. 5. A MV grid with grid monitoring devices

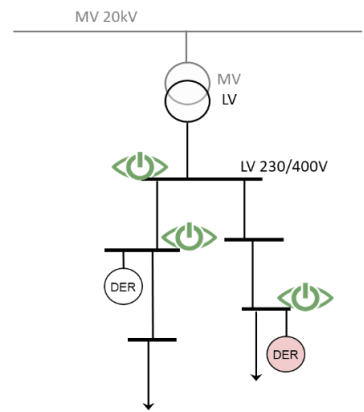

Fig. 6. A representative LV grid with grid monitoring devices

The results and discussions have been based on two use cases. *Case 1* represents the case where the grid is operated optimally with the combined formulation of model-based MV and model-less LV grids. Finally, *Case 2* denotes the case with a separate fast and slow flexibility services estimation at the P-SS. The distinction between the fast and the slow services are decided by the ramping limit constraints, as explained in the previous section.

### A. Results

*1) Combined operation of model-based MV grid and model-less LV grid (Case 1):* The combined operation of the MV and LV grids ensures the optimal operation of the grid with the central platform hosting the grid aware algorithm. The formulation reduces the technical losses, security violation costs and the deviation of P-SS power flows from the scheduled flows.

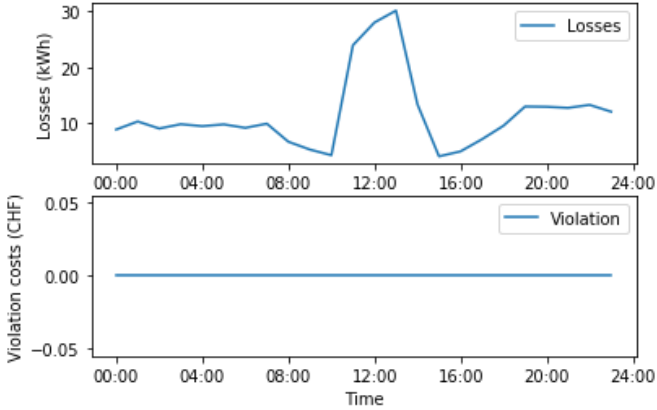

Fig. 7. Technical losses (kWh) and security violation costs (CHF) for the combined operation case.

Fig. 7 shows the technical losses in kWh and security violation costs in CHF (calculated with the violations from the limits being multiplied by 100 CHF) for a typical day. The losses are higher during the day when the PV output is high due to the flow of reactive power, as the PV inverters are assumed to be in power factor control modes. This further shows that there are no security violations (voltage or line flows) in the MV grid when the LV DER flexibilities are activated at the current loading level of the network.

*2) Aggregated flexibility estimation at P-SS with fast and slow services distinction (Case 2):* This case represents the flexibility estimation from the distribution grids to the TSO with separate P-Q flexibility curves for fast and slow services. This P-Q curve takes into consideration the operational limits of node voltages and line flows for both the MV and LV grids. This provides the TSO and/or DSO an estimation of available flexibility of fast and/or slow should the need arises. It is assumed, here, that the slow flexibility services have ramping limits less than 4 kW/hr and the fast flexibility services have ramping limits above 4 kW/hr. The methodology illustrates the quantity of a fast or a slow service, from resources connected at the LV level, available at the P-SS, if requested by the TSO, whilst ensuring minimum voltage and flow violations for the MV distribution grid.

### B. Discussions

The results from the combined MV and LV grid optimisation show that it is possible to achieve the optimal operation of the distribution grid with model-based MV and model-less LV grids whilst realising flexibility potential from individual DERs (PVs, EVs) connected at the LV level. Fig. 7 shows the flexibility potential of DERs can be realised by this methodology and limited by the grid security limits (voltage and line flows). The methodology is complimented by the presence of a central infrastructure facilitated by accurate and granular grid monitoring data.

Fig. 8 shows the separate fast and slow P-Q flexibility curves at (a) high EV and no PV output (11pm), (b) limited PV and EV output (6pm), (c) high PV and low EV output (1pm) of a typical day in September 2020 to differentiate between the fast and slow flexibilities available from the distribution grids. This provides the TSO an estimate of how much flexibility potential can be procured or activated should the requirement arise. For each of the scenarios (for different levels of EV and PV output), it is observed that the fast and slow flexibility curves can be estimated separately. The result is beneficial for both the TSO leader and DSO leader coordination schemes. In the TSO leader scheme, the aggregated estimation at P-SS means that both fast and slow services are pre-qualified by the DSO by this methodology. In the DSO leader scheme, the fast or slow flexibilities are aggregated at P-SS only after the DSO has utilised the flexibilities for maintaining its own grid security.

The results infer that this methodology, proposed here, along with its model-less LV grid approach, is particularly beneficial to realise the optimal operation and flexibility potential from the LV grid DERs whilst maintaining the secure operation of the distribution grid. In the age of increasing decentralised and intermittent generations, the methodology of flexibility estimation of separate fast and slow flexibility services to the TSO from small-scale LV DERs will provide the grid operators (TSO and DSO) a roadmap for operational planning.

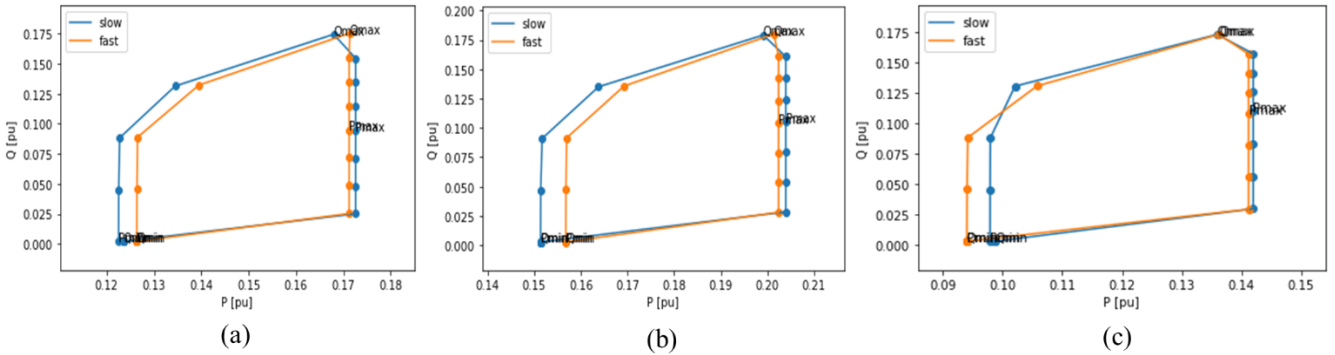

Fig. 8. Aggregated fast and slow flexibility estimation at P-SS at (a) high EV and no PV, (b) limited PV and EV output, (c) high PV and low EV.

## V. CONCLUSION

This paper proposed a methodology of provision and activation of flexibilities connected at the LV grids. To enable the optimal, secure and reliable operation of the grid this paper developed a combined optimal power flow for the MV and LV grids. The methodology based on a model-based MV grid and sensitivity coefficients-based LV grids ensures that there is no requirement of an accurate and up-to-date LV grid topology. The paper, further, proposed a methodology for provision of fast and slow flexibility services from the resources connected at the LV grids.

The results of this paper provide valuable insights to the continuously evolving distribution system operation. The DSO will be able to reduce their costs of operation while ensuring security and quality of supply. This approach enables the DSOs to improve flexibility potential and improve the hosting capacity of intermittent renewables. It will facilitate the demand response programs through EVs, TCLs, solar PVs, ESS. Furthermore, as TSOs procure more resources from the DSO, this methodology will enable the TSO and the DSOs to realise the flexibility aggregation of fast and slow services to be provided for the TSO ancillary and balancing services. As a future step, for the sake of completion, the provision of even faster primary responses will be included in the study provided that sufficient flexibility capacity is available in the LV grids. In summary, this work will provide a flexibility roadmap to the DSOs and the TSO to facilitate a decentralised, decarbonised and digitalised system.